# *Large spontaneous exchange bias in a weak ferromagnet $Pb_6Ni_9(TeO_6)_5$*

**B. Koteswararao,[1] Tanmoy Chakrabarty,[2] Tathamay Basu,[2,3] Binoy Krishna Hazra,[1] P. V. Srinivasarao,[4] P. L. Paulose,[2] S. Srinath[1]**

[1]*School of Physics, University of Hyderabad, Hyderabad-500046, India*
[2]*Tata Institute of Fundamental Research, Homi Bhabha Road, Colaba, Mumbai- 400005, India*
[3]*Experimental Physics V, Center for Electronic Correlations and Magnetism, University of Augsburg, Augsburg, D- 86159, Germany*
[4]*Department of Physics, Acharya Nagarjuna University, Nagarjuna Nagar 522 510, India*

We report the magnetic and dielectric behavior of $Pb_6Ni_9(TeO_6)_5$, a new compound comprising the honeycomb-like layers of *S*=1 spins, through detailed structural, magnetic and dielectric investigation. An antiferromagnetic-type transition at 25 K ($T_N$) with weak-ferromagnetic behavior is revealed. Interestingly, a large value of coercive field of 1.32 T at 2 K is observed. The isothermal magnetization after zero-field-cooled condition, it exhibits the presence of large spontaneous exchange bias (SEB) with a magnitude of 0.19 T at 2 K; which is rare in single bulk materials, especially without external doping. The value of |$H_{EB}$| further enhances to 0.24 T under 16 T field-cooled condition, confirming the presence of large exchange bias in the material. In addition, the dielectric constant shows an anomaly at the onset of $T_N$, indicating the presence of magnetodielectric coupling.

**Introduction**

Conventional exchange bias (CEB) effect is a magnetic phenomenon observed by a shift along the magnetization and magnetic field axis in the magnetic hysteresis loop when the material is cooled in a magnetic field ($H$)[1,2]. Initially, the CEB effect has been observed in a system composed of antiferromagnetic (AFM) and ferromagnetic (FM) constituents where the Curie temperature of the FM substance is greater than that of the Neel temperature corresponding to the AFM substance. The uncompensated AFM spins at the interface due to the coupling of FM spins adjacent to AFM spins, gives rise to the unidirectional anisotropy causes the displacement of the magnetic hysteresis loop. Later, CEB effect has been also noticed in various combinations between FM, AFM, canted AFM, ferrimagnetic (FIM), and spin glass (SG) magnetic components[3-6].

 On the contrary to the CEB appeared due to the field-cooling process, the spontaneous exchange bias (SEB) has been observed in certain magnetic systems without the assistance of static *H* cooling. As there is no requirement of external *H* for generating the unidirectional anisotropy, the SEB effect will have greater applications in the electric field controlled exchange bias (EB) devices. Recently, this unusual zero-field-cooled (ZFC) EB or large SEB has been observed in a few systems; $BiFeO_3$-$Bi_2Fe_4O_9$ nanocomposite, $NiMnIn_{13}$, $Mn_2PtGa$, $La_{1.5}Sr_{0.5}MnCoO_6$, and $Co_{0.75}Cu_{0.25}Cr_2O_4$ polycrystalline samples[7-12]. The presence of multiple phases in these materials could be the reason for the observation of spontaneous unidirectional anisotropy. For instance, the formation of SG phase at the interface of multiferroic $BiFeO_3$-$Bi_2Fe_4O_9$ nanocomposite was found to be responsible to the observation of SEB effect; however the CEB field value was found to be much larger than the SEB field value[7]. In case of Mn$_2$PtGa, the SEB field value, where the effect developed due to FM clusters embedded in FIM ordered matrix, was interestingly similar to its CEB value[8]. However, the origin of this effect is still under debate and varies system to system. To further understand the SEB behavior, a few more new materials exhibiting SEB effect need to be explored. Moreover, a few unique magnetic materials only exhibit this effect with large value of SEB field and especially rare in case of single bulk material. In this publication, we report the magnetic properties of a new $S$=1 spin system Pb$_6$Ni$_9$(TeO$_6$)$_5$ which exhibits a transition at 25 K with weak ferromagnetic (WFM) behavior. Interestingly, a large value of coercive field about 1.32 T and SEB field of about 0.19 T at 2 K are observed when the sample is cooled under ZFC condition. The value of such high SEB is rare, especially in single bulk material without any external doping. There is no previous report of magnetic and dielectric behavior of this interesting compound.

**Results and Discussion**

The compound has been characterized by XRD analysis (shown in Fig. 1(a), a more clear view of this figure is shown in Supplementary Information). We have performed the Rietveld refinement analysis to check the phase purity of the sample[13]. The compound forms in desired structure (space group P6$_3$22)[14], we did not find any impurity phase within the resolution of the instrument. The lattice parameters estimated are about $a=b$=10.288(5) Å, and $c$=13.533(5) Å, and $\alpha=\beta$=90°, $\gamma$=120°, which nearly agree with the reported[14] values of $a=b$=10.2579(10) Å, and $c$=13.554(5) Å. The obtained residual parameters from the Rietveld refinement are $R_p \approx$24.3%, $R_{wp} \approx$22.3%, $R_{exp} \approx$8.55%, and $\chi^2 \approx$6.83, respectively. These values are slightly bigger compared to the reported values of the single crystal data, which could be due to site-exchange between Ni and Te atoms. In addition, there are not much differences between our values of bond-lengths and bond-angles to those of earlier reported values. The obtained atomic positions of different atoms of this compound are mentioned in Table SI in Supplementary Information.

The crystal structure is depicted in Fig. 1(b). The compound forms in hexagonal structure which comprises NiO$_6$, TeO$_6$ octahedral units and Pb atoms. The Ni atoms in the *ab*-plane form a honeycomb-like lattice (see Fig. 1 (d)-(f)). These layers are not very well separated due to which the feature might not attribute to 2D magnetism, unlike to the perfect $S$=1 honeycomb magnetic material BaNi$_2$V$_2$O$_8$[15]. The bond-angles of Ni-O-Ni in the $z$=0.25 layer (which indeed forms a 1/6$^{th}$ depleted Honeycomb layer), built by *6h* and *2b* sites of Ni atoms, are in the range of 90°-92° (see Table SII in Supplementary Information). According to Goodenough predictions[16], the magnetic coupling favors ferromagnetic (FM)-type. On the other hand, the bond-angle between the Ni atoms in $z$=0.25 layer and Ni1 atom *via* O3 is about 122.2°, which usually allow AFM interactions[16]. In addition, these honeycomb-like layers (1/6$^{th}$ depleted honeycomb layer with additional triangular-like interactions, see Fig. 1(e)) might allow the competition between AFM and FM couplings which might lead to the unusual magnetism.

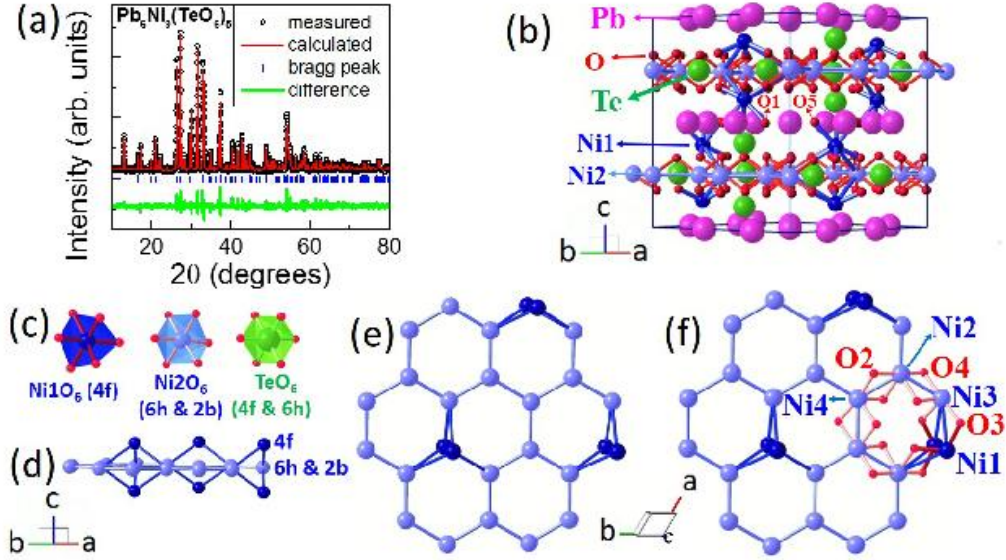

**Figure 1.** (a) Rietveld refinement on the XRD of the powder samples. (b) Crystal structure of $Pb_6Ni_9(TeO_6)_5$. (c) The distorted $NiO_6$ (*4f*, *6h*, and *2b*) and $TeO_6$ (*4f* and *6h*) octahedral environments in the structure. (d), (e) and (f) show the honeycomb-like layer in *ab*-plane filled with $Ni^{2+}$ ions. The layer is formed by four $Ni^{2+}$ ions (Ni1, Ni2, Ni3, and Ni4) and these ions are interacted each other through $O^{2-}$ ions.

The dc-magnetic susceptibility $\chi$ (=$M/H$) is measured as a function of $T$ in $H$ of 0.5 T (see Fig. 2). At high-$T$, the data follow the Curie-Weiss behavior. Inverse-$\chi(T)$ is fitted to $(T-\theta_{CW})/C$ and the obtained effective magnetic moment of $Ni^{2+}$ is 3.18 $\mu_B$, consistent with the value obtained from other Ni-based magnets[17]. The Curie-Weiss temperature ($\theta_{CW}$) is obtained to be -30 K, suggesting that the dominant magnetic couplings are of AFM at high-$T$. At low-$T$, there is a sudden upturn seen at 25 K ($\approx T_N$) with a change in the slope at about 20 K. There is a clear bifurcation between the ZFC and FC data, probably due to the appearance of anisotropic field. The dielectric constant $\varepsilon'(T)$ is measured in the frequency range 1- 100 kHz. The dielectric features does not show any frequency dependent behavior, therefore one frequency data is only shown in the inset of Fig. 2. $\varepsilon'(T)$ falls steeply from high-$T$ and shows an upturn below 100 K, followed by a clear feature below $T_N$, indicating the presence of magnetodielectric (MDE) coupling (the cross-coupling between spin and lattice). We have not observed any significant changes of this dielectric feature under application of $H$, could be due to weak MDE coupling. The feature at high $T$ (upturn below 100 K) may arise due to the presence of AFM couplings (short-range correlation above ordering). Similar upturn in the $\varepsilon'(T)$ has also been observed in many multiferroic and/or magneto-electric oxides due to magnetic correlations[18-20]. We do not want to elaborate this further which is not the aim of this manuscript.

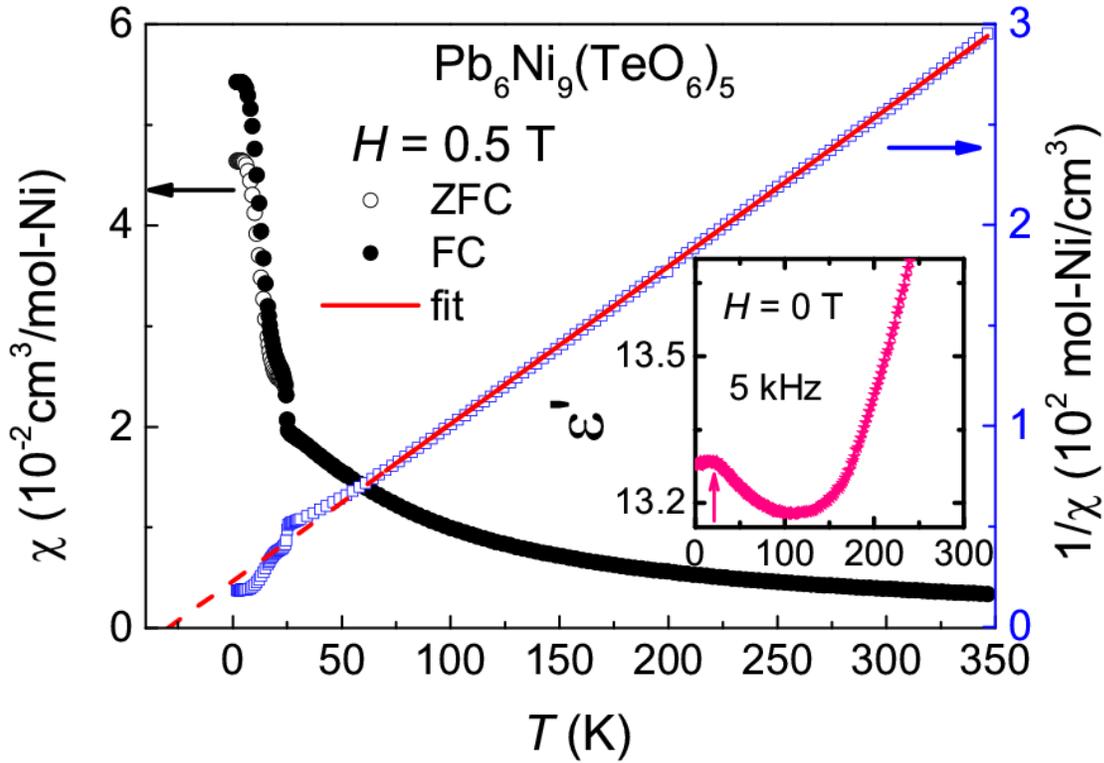

**Figure 2.** $\chi(T)$ measured in 0.5 T. The solid line is a fit to $1/\chi$ data, while the dashed line represents the extrapolation. Inset shows the $\varepsilon'(T)$ measured at frequency 5 kHz and 0 T.

The $\chi(T)$ is plotted for different fields 0.002 T, 2 T, 5 T, and 15 T (see Fig. 3(a)-(d)). At low-fields, the bifurcation of ZFC and FC data is large. The bifurcation reduces with increasing $H$. A cusp like behavior is seen at $H$ above 2 T indicating that the system is that of AFM-type assisted with some weak ferromagnetic moments. Moreover, the bifurcation of ZFC and FC disappears at 15 T (see Fig. 3(d)), which assures that the magnetic field of 15 T saturated the weak ferromagnetic moment or uncompensated moment. Further, the *ac*-magnetic susceptibility ($\chi_{ac}(T)$) is performed in the frequencies of 11 Hz and 111 Hz. The real part of $\chi_{ac}$ ($\chi'_{ac}(T)$ shows a peak at 25 K (see Fig. 3(a)) and there is no frequency dependent behavior seen across the transition, suggesting the absence of glassy nature of magnetic moments; no feature is observed in imaginary part of *ac*-susceptibility ($\chi''_{ac}(T)$) which confirms this conclusion. The observed results from the magnetic susceptibility studies suggest that $Pb_6Ni_9(TeO_6)_5$ exhibits a WFM behavior, which is possibly due to the presence of Dzyaloshinskii–Moriya (DM) anisotropic interactions presented in this magnetic material[21].

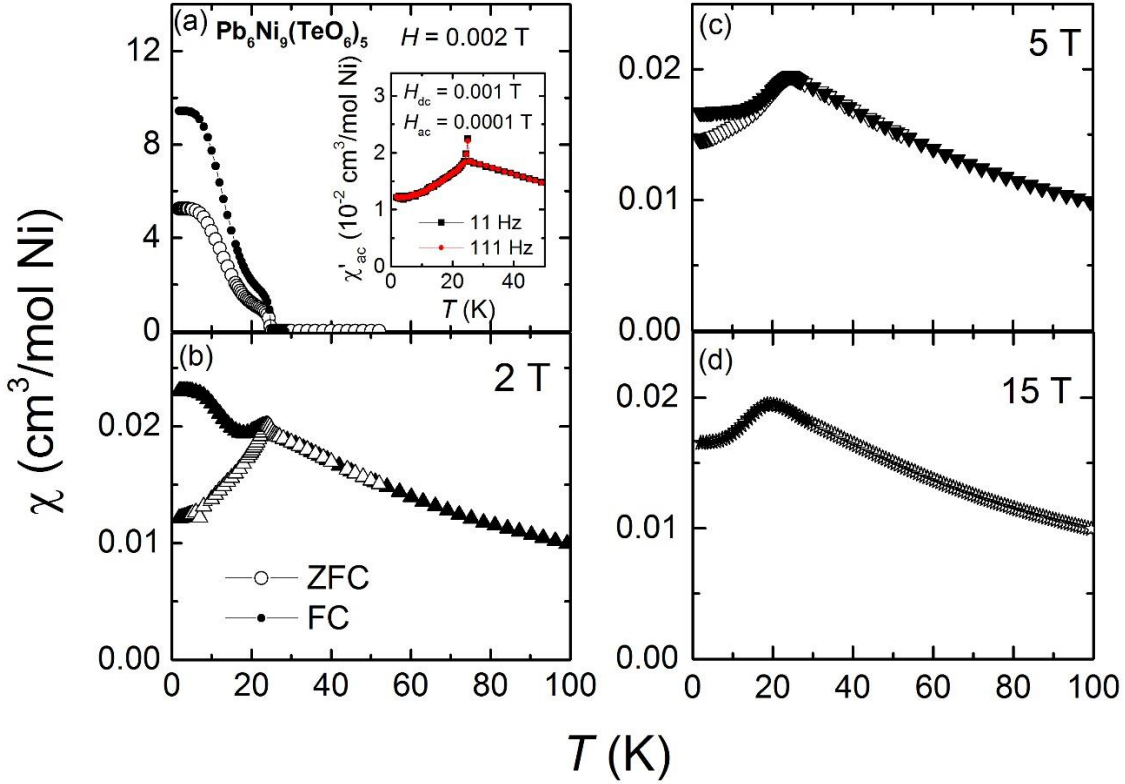

**Figure 3.** ZFC and FC $\chi(T)$ at different fields. Open symbols indicate the ZFC data, while the filled circles indicate field cooled (FC) data. The inset of (a) shows the plot of $\chi'_{ac}(T)$ at the frequencies of 11 Hz and 111 Hz.

The isothermal magnetization $M(H)$ are measured at 2 K in different ZFC and FC conditions (see Fig. 4(a)). In order to get a major loop, the data was collected until the maximum $H$ of 16 T, which is higher than the anisotropic field; this is an essential condition to confirm the intrinsic exchange bias present in this compound. At 2 K, the data show a typical WFM behavior. The derivative of $M$ with respect to $H$, *i.e.*, $dM/dH$ versus $H$ plot is shown in Fig. 4(b). It can be clearly seen in Fig. 4(a) & 4(b) that the bifurcation between the data of $H$-increasing and $H$-decreasing disappears above 13 T, indicating that our $M(H)$ measured until 16 T is a major loop[22]. In addition, a change in the slope is observed at 15 T, indicating the spin-flop transition. Interestingly, even under ZFC condition, it shows a highly asymmetric loop with a clear shift in $M$ and $H$-axis: a typical behavior of an exchange bias (EB) (see inset (i) of Fig. 4(a)). The values of coercive fields at positive-$H$ and negative-$H$ axis are found to be $H_{c+} \approx +1.13$ T and $H_{c-} \approx -1.51$ T, respectively. Large value of coercive $H_c = (H_{c+} - H_{c-})/2 \approx 1.32$ T is observed in this WFM system. The value of ZFC EB or SEB field at 2 K is estimated to be about $|H_{EB}| = |-(H_{c+} + H_{c-})/2| \approx 0.19$ T. The EB along $M$-axis ($M_{EB}$) is also observed and the value is found to be about 0.03 $\mu_B/Ni^{2+}$.

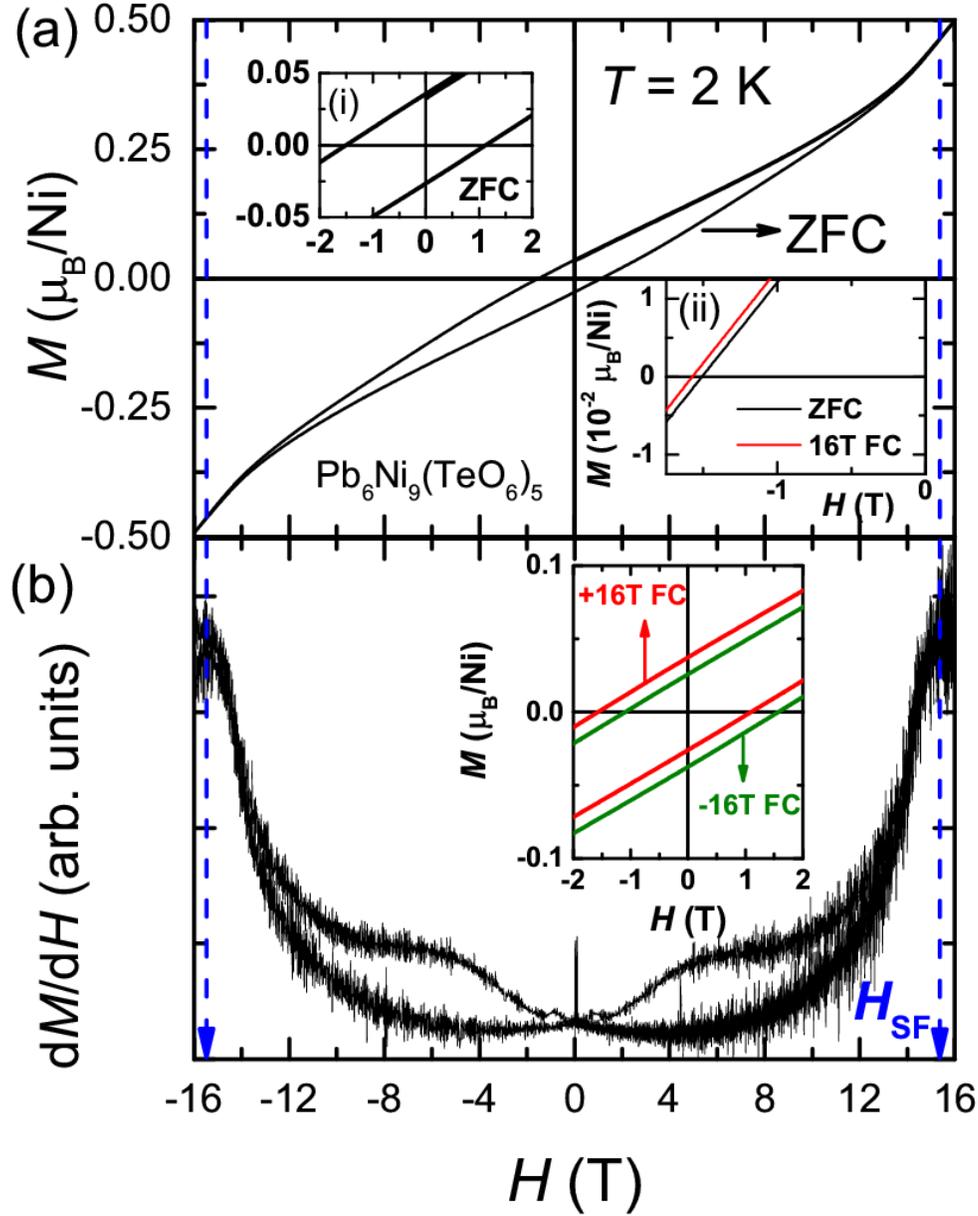

**Figure 4.** (a) ZFC $M(H)$ at 2 K. Inset (i) shows the zoomed part of ZFC data at 2 K. Inset (ii) shows ZFC and +16 T FC $M(H)$ at 2 K. (b) plots $dM/dH$ vs. $H$. Inset of (b) shows the $M(H)$ plots obtained after cooling in $H$ of +16 T and -16 T.

Recently, similar kind of SEB has been observed in a few magnetic materials, however, such a high value of SEB is rare[8,9,10,12,23,24], specially in a single bulk material without any external doping. Similar large value of SEB has been reported in a few doped bulk materials only ($La_{1.5}Sr_{0.5}CoMnO_6$, $Sm_{1.5}Ca_{0.5}CoMnO_6$, specific doped Heusler Alloy system like Mn–Pt–Ga or Ni–Mn–Ga) [10,12,23,24].

We also measured $M(H)$ loop at 2 K after 16 T FC and -16 T FC conditions. As shown in the inset (ii) of Fig. 4(a), a shift between ZFC and 16 T FC $M(H)$ is observed, claiming the appearance of conventional exchange bias (CEB) in this material. The value of $|H_{EB}|$ is increased to 0.24 T for the data measured under +16 T FC condition. In addition, there is also a clear shift between the *M*(*H*) isotherms after +16 T FC and -16 T FC conditions, respectively (see inset of Fig. 4(b)). Therefore, the observation of asymmetric nature in the ZFC *M*(*H*) and the further enhancement of this effect in FC *M*(*H*) data is unambiguously claiming that the presence of EB or unidirectional anisotropy in the prepared sample.

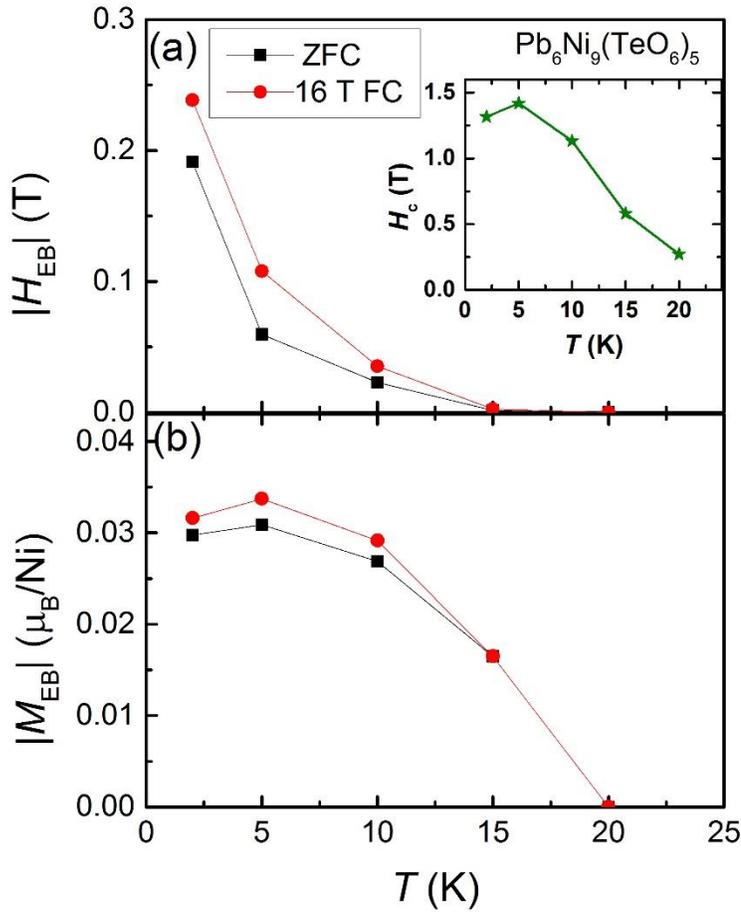

**Figure 5.** (a) and (b). The values of |$H_{EB}$| and |$M_{EB}$| vs. *T*. Inset of (a) plots $H_c$ vs. *T*.

To further understand this effect, we have measured *M*(*H*) at different temperatures after cooling the sample under ZFC and 16 T FC conditions. As *T* increases, the values of both SEB and CEB fields (|$H_{EB}$|) and magnetization (|$M_{EB}$|) decrease (see Fig. 5(a) & (b)). As shown in the inset of Fig. 5(a), the value of $H_C$ increases to 1.5 T at 5 K, which is comparable with the permanent magnets 2.13 T for $Nd_2Fe_{14}B$, 0.6 T for $SmCo_5$ at room temperature[25], and 2.7 T for organic-based Mn-chain magnet at 2 K[26]. The value of $H_c$ further decreases with increasing *T*.

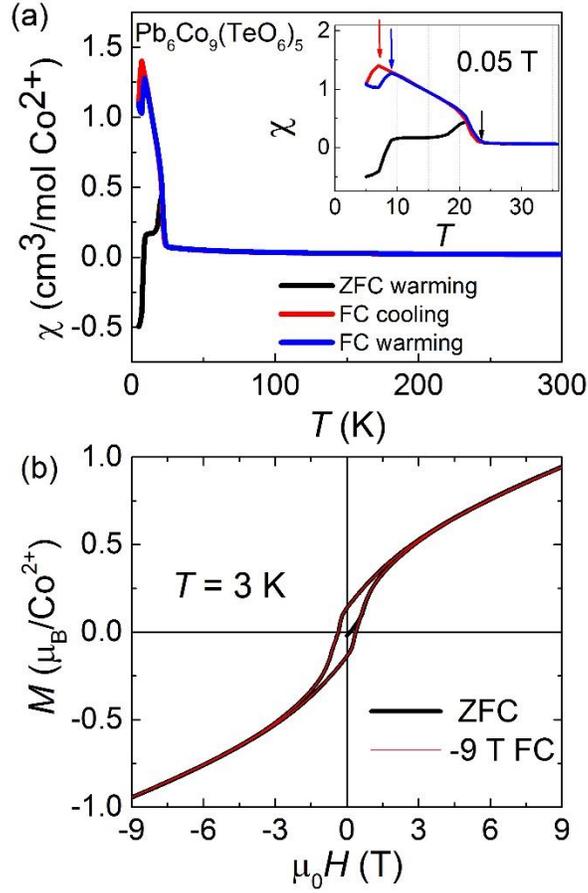

Fig. 6 χ(*T*) measured at 0.05 T on the polycrystalline samples of $Pb_6Co_9(TeO_6)_5$. Inset shows the magnetic transitions at low-*T*. (b) *M*(*H*) at 3 K after ZFC and -9 T FC.

This unusual, large EB behavior in $Pb_6Ni_9(TeO_6)_5$, can be understood as follows. The complex magnetism usually comes from a different variety of magnetic interactions present (isotropic and anisotropic) and the competition between them. From the structural point of view, $Pb_6Ni_9(TeO_6)_5$ has a honeycomb-like layers with bond-angles of Ni-O-Ni in the range varying from 90º to 120º, which favors both FM and AFM couplings, which is also evidenced in χ(*T*). The observation of WFM behavior suggests the presence of DM anisotropic interactions[21]. It has been also theoretically predicted that the importance of DM anisotropic interactions in attributing EB[27].

On the other hand, $Pb_6Co_9(TeO_6)_5$ exhibit an identical crystal structure to that of present system $Pb_6Ni_9(TeO_6)_5$[28]. Similar to $Pb_6Ni_9(TeO_6)_5$, the compound $Pb_6Co_9(TeO_6)_5$ also exhibit AFM correlations with θcw≈-35 K at high-*T* and WFM behavior below its transition at 25 K (χ(*T*) plot is not shown). *M*(*H*) plot at 3 K evidence the WFM behavior (see inset of Fig. 5(b). However, it does not exhibit any EB under ZFC and FC conditions. The variation in the magnetic property of Co-magnet might be due to the larger size of $Co^{2+}$ (88 pm) compared to that of $Te^{6+}$ (70 pm) in the octahedral environments, which usually does not allow the site-exchange. Since the difference in the ionic sizes of $Ni^{2+}$ and $Te^{6+}$ (82 pm and 70 pm) is relatively small, we thus consider the possibility of a little site-exchange between $Ni^{2+}$ and $Te^{6+}$ ions in $Pb_6Ni_9(TeO_6)_5$. This anti-site disorder might induces a microstructural magnetic phase, which is embedded with the WFM host. Recently, it has been also reported that the anti-site disorder in and $La_{1.5}Sr_{0.5}CoMnO_6$[29] and $Mn_2PtGa$[9,23] materials which developed the FM clusters/microstructure. The coupling of these embedded secondary phases with the host leads to the SEB. Hence, we believe that the SEB in $Pb_6Ni_9(TeO_6)_5$ could be due to the coupling between WFM host resulted from DM anisotropic interactions and the embedded microstructure formed due to the site-exchange. However, the detailed measurements on high-resolution synchrotron, neutron diffraction and electron-spin-resonance spectroscopy would help in confirming the local site-exchange and also understanding the nature of anisotropic couplings.

**Conclusion**

In conclusion, we have observed a large value of SEB field of 0.19 T and coercive field of 1.32 T at 2 K in the sample $Pb_6Ni_9(TeO_6)_5$. Such high value of SEB (without application of external *H*) in a single material is rare, specifically, without any external doping. This EB value is enhanced further to 0.24 T when the sample is cooled under 16 T. The coupling between the WFM host and the microstructure that probably arises from a small percentage of site-exchange of $Te^{6+}$ and $Ni^{2+}$ ions might play a role in having such a large SEB and/or unidirectional anisotropy in this magnetic system. It is worthwhile to explore more about this WFM system through both experimental and theoretical point of view for designing the future materials with large values of EB at room temperature.

**Methods**

The compounds were synthesized in solid-state-reaction method. The high purity (>99.95%) chemicals of PbO, NiO or $Co_3O_4$, and $TeO_2$ were mixed in stoichiometric ratio, grounded thoroughly and fired at 600°C, 700°C, 750°C, and 820°C, respectively. Each firing was carried out for 12 hours with several intermediate grinding. X-ray diffraction (XRD) was performed in the 2θ range from 8° to 90° using Rigaku X-ray diffractometer. Temperature (*T*) and magnetic field (*H*) dependent magnetization (*M*) were carried out in the *T* range 2-350 K and up to ±16 T magnetic field using vibrating sample magnetometer (VSM, Quantum Design). The large magnetic fields are purposefully applied for this material to satisfy the major *M*(*H*) loop condition and the presence of exchange bias in this material, according the Ref. [22]. Dielectric constant (*ε*) measurement were performed using LCR meter (4980A, Agilent) which is integrated to Physical Properties Measurement System (PPMS, Quantum Design) to control *T* and *H*.

**References**

1. Meiklejohn, W. H. & Bean, C. P. New magnetic anisotropy. *Phys. Rev.* **102,** 1413-1414 (1956).
2. Meiklejohn, W. H. & Bean, C. P. New magnetic anisotropy. *Phys. Rev.* **105**, 904-913 (1957).
3. Nogués, J. & Schuller, I. K. Exchange bias. *J. Magn. Magn. Mater.* **192**, 203-232 (1999).
4. Nogués, J., Sort. J., Langlais, V., Skumryev, V., Suriñach, S., Muñoz, J. S. & Baró, M. D. Exchange bias in nanostructures. *Phys. Rep.* **422,** 65-117 (2005).
5. Berkowitz, A. E. & Takano, K. Exchange bias - a review. *J. Magn. Magn. Mater.* **200**, 552-570 (1999).

6. Giri, S., Patra, M. & Majumdar, S. Exchange bias effect in alloys and compounds. *J. Phys.: Condens. Matter* **23,** 073201 (2011).
7. Maity, T., Goswami, S., Bhattacharya, D. & Roy, S. Superspin Glass Mediated Giant Spontaneous Exchange Bias in a Nanocomposite of $BiFeO_3$-$Bi_2Fe_4O_9$. *Phys. Rev. Lett.* **110**, 107201 (2013).
8. Wang, B. M. *et al.* Large Exchange Bias after Zero-Field Cooling from an Unmagnetized State. *Phys. Rev. Lett.* **106**, 077203 (2011).
9. Nayak, A. K. *et al*. Large Zero-Field Cooled Exchange-Bias in Bulk $Mn_2PtGa$. *Phys. Rev. Lett.* **110**, 127204 (2013).
10. Murthy, J. K. & Venimadhav, A. Giant zero field cooled spontaneous exchange bias effect in phase separated $La_{1.5}Sr_{0.5}CoMnO_6$. *Appl. Phys. Lett.* **103**, 252410 (2013).
11. Wang, L. G. *et al*. Negative magnetization and zero-field cooled exchange bias effect in $Co_{0.8}Cu_{0.2}Cr_2O_4$ ceramics. *Appl. Phy. Lett.* **107**, 152406 (2015).
12. Giri, S. K., Sahoo, R. C., Dasgupta, P., Poddar, A. & Nath, T. K. Giant spontaneous exchange bias effect in $Sm_{1.5}Ca_{0.5}CoMnO_6$ perovskite. *J. Phys. D: Appl. Phys.*, **49**, 165002 (2016).
13. Juan Rodruez-Carvajal. Recent advances in magnetic structure determination by neutron powder diffraction. *Physica B* **192**, 55 (1993).
14. Wedel, B. *et al.* Verknuepfung von $(TeO_6)_6$ und $(TeO_6)_3$ $(NiO_6)_3$ sechringen durch $TeNiO_9$-Oktaederdoppel in $Pb_3Ni_{4.5}Te_{2.5}O_{15}$ *Z. Naturforsch. B: Chem. Sci.*,**53**, 527-531 (1998).
15. Rogado, N., Huang, Q., Lynn, J. W., Ramirez, A. P., Huse, D. & Cava, R. J. $BaNi_2V_2O_8$: A two-dimensional honeycomb antiferromagnet. *Phys. Rev. B* **65**, 144443 (2002).
16. Goodenough, J. B. Theory of the Role of Covalence in the Perovskite-Type Manganites [La, M(II)] $MnO_3$. *Phys. Rev.*, **100**, 564 (1955).
17. Cheng, J. G., Li, G., Balicas, L., Zhou, J. S., Goodenough, J. B., Xu, C., & Zhou, H. D. High-Pressure Sequence of $Ba_3NiSb_2O_9$ Structural Phases: New $S = 1$ Quantum Spin Liquids Based on $Ni^{2+}$. *Phys. Rev. Lett.* **107**, 197204 (2011).
18. Law, J. M., Reuvekamp, P., Glaum, R., Lee, C., Kang, J., Whangbo, M. H. & Kremer, R. K. Kremer. Quasi-one-dimensional antiferromagnetism and multiferroicity in $CuCrO_4$. *Phys. Rev. B* **84**, 014426 (2011).
19. Koteswararao, B., Panda, S. K., Kumar, R., Yoo, K., Mahajan, A. V., Dasgupta, I., Chen, B. H., Kim, K. H. & Chou, F. C. Observation of $S = 1/2$ quasi-1D magnetic and magneto-dielectric behavior in a cubic $SrCuTe_2O_6$. *J. Phy: Cond. Matt* **27**, 426001 (2015).
20. Koteswararao, B., Yoo, K., Chou, F. C. & Kim, K. H. Observation of magnetoelectric effects in a $S = 1/2$ frustrated spin chain magnet $SrCuTe_2O_6$. *APL Materials* **4**, 036101 (2016).
21. Moriya, T. Anisotropic Superexchange Interaction and Weak Ferromagnetism. *Phys. Rev.* **120**, 91 (1960).
22. Harres, A., Mikhov, M., Skumryev, V., de Andrade, A. M. H., Schmidt, J. E., Geshev, J. Criteria for saturated magnetization loop. *J. Mag. Mag. Mater.* **402**, 76 (2016).
23. Nayak, A. K. *et al.* Design of compensated ferrimagnetic Heusler alloys for giant tunable exchange bias. *Nat. Mater.* **14**, 679 (2015).
24. Tian, F. *et al.* Giant spontaneous exchange bias triggered by crossover of superspin glass in Sb-doped $Ni_{50}Mn_{38}Ga_{12}$ Heusler alloys. *Sci. Rep.* **6**, 30801 (2016).
25. Long, G. J. & Grandjean, F. "*Supermagnets, Hard Magnetic Materials*", Kluwer academic publications, Dordrecht, pp 1-6. (1991).

26. Rittenberg, D. K., Sugiura, K., Sakata, Y., Mikami, S., Epstein, A. J. & Miller, J. S. Large Coercivity and High Remanent Magnetization Organic-Based Magnets. *Adv. Mater.*, **12**, 126 (2000).
27. Dong, S., Yamauchi, K., Yunoki, S., Yu, R., Liang, S., Moreo, A., Liu, J. M., Picozzi, S. & Dagotto, E. Exchange Bias Driven by the Dzyaloshinskii-Moriya Interaction and Ferroelectric Polarization at G-Type Antiferromagnetic Perovskite Interfaces. *Phys. Rev. Lett.* **130**, 127201 (2009).
28. Christine, A. & Matthias, W. $Pb_6Co_9(TeO_6)_5$. *Acta Cryst. E* **68**, i71 (2012).
29. Murthy, J. K., Chandrasekhar, K. D., Wu, H. C., Yang, H. D., Lin, J. Y. & Venimadhav, A. Antisite disorder driven spontaneous exchange bias effect in $La_{2-x}Sr_xCoMnO_6$ (0<x<1). *J. Phy.:Cond. Matter* **28,** 086003 (2016).

**Acknowledgments**

BK thanks DST for INSPIRE faculty award-2014.

**Author Contributions**
BKR synthesized the samples. TC, TB, BKH has measured the experimental data. All authors analyzed the results, write the manuscript. BKR and SS reviewed the manuscript.

**Additional Information**
Supplementary information accompanies this paper.
Competing financial interests: The authors declare no competing financial interests.

**Figure Captions:**

**Figure 1.**

(a) Rietveld refinement on the XRD of the powder samples. (b) Crystal structure of $Pb_6Ni_9(TeO_6)_5$. (c) The distorted $NiO_6$ (*4f*, *6h*, and *2b*) and $TeO_6$ (*4f* and *6h*) octahedral environments in the structure. (d), (e) and (f) show the honeycomb-like layer in *ab*-plane filled with $Ni^{2+}$ ions. The layer is formed by four $Ni^{2+}$ ions (Ni1, Ni2, Ni3, and Ni4) and these ions are interacted each other through $O^{2-}$ ions.

**Figure 2.**

$\chi(T)$ measured in 0.5 T. The solid line is a fit to $1/\chi$ data, while the dashed line represents the extrapolation. Inset shows the $\varepsilon'(T)$ measured at frequency 5 kHz and 0 T.

**Figure 3.**

ZFC and FC $\chi(T)$ at different fields. Open symbols indicate the ZFC data, while the filled circles indicate field cooled (FC) data. The inset of (a) shows the plot of $\chi'_{ac}(T)$ at the frequencies of 11 Hz and 111 Hz.

**Figure 4.**

(a) ZFC $M(H)$ at 2 K. Inset (i) shows the zoomed part of ZFC data at 2 K. Inset (ii) shows ZFC and +16 T FC $M(H)$ at 2 K. (b) plots $dM/dH$ vs. $H$. Inset of (b) shows the $M(H)$ plots obtained after cooling in $H$ of +16 T and -16 T.

**Figure 5.**

(a) and (b). The values of $|H_{EB}|$ and $|M_{EB}|$ vs. $T$. Inset of (a) plots $H_c$ vs. $T$.

**Figure 6.**

$\chi(T)$ measured at 0.05 T on the polycrystalline samples of $Pb_6Co_9(TeO_6)_5$. Inset shows the magnetic transitions at low-$T$. (b) $M(H)$ at 3 K after ZFC and -9 T FC.

# Supplementary Information: *Large spontaneous exchange bias in a weak ferromagnet Pb$_6$Ni$_9$(TeO$_6$)$_5$*

B. Koteswararao,[1] Tanmoy Chakrabarty,[2] Tathamay Basu,[2,3] Binoy Krishna Hazra,[1] P. V. Srinivasarao,[4] P. L. Paulose,[2] S. Srinath[1]

[1]*School of Physics, University of Hyderabad, Hyderabad-500 046, India*
[2]*Tata Institute of Fundamental Research, Homi Bhabha Road, Colaba, Mumbai- 400005, India*
[3]*Experimental Physics V, Center for Electronic Correlations and Magnetism, University of Augsburg, Augsburg, D- 86159, Germany*
[4]*Department of Physics, Acharya Nagarjuna University, Nagarjuna Nagar 522 510, India*

**Rietveld refinement of powder X-ray diffraction**

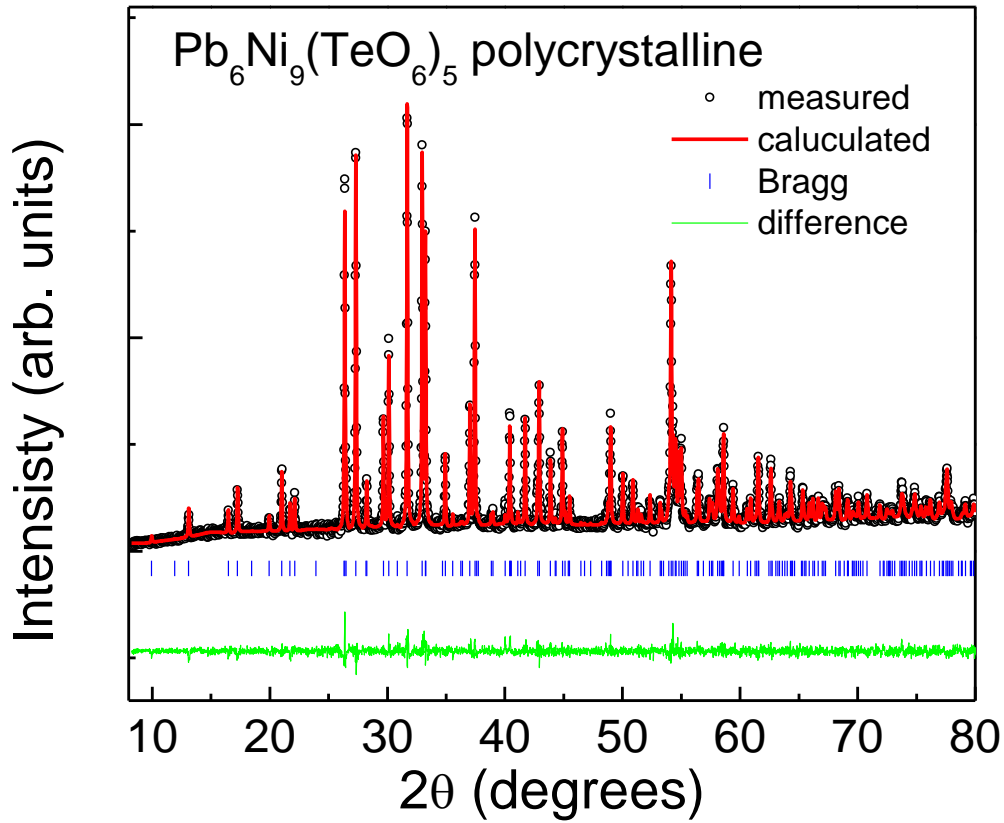

**Figure S1**: Rietveld refinement on the powder XRD of Pb$_6$Ni$_9$(TeO$_6$)$_5$ samples.

**Table S I: The details of atomic positions of various atoms in Pb$_6$Ni$_9$(TeO$_6$)$_5$**

| Atom | Wyckoff position | x/a | Y | Z | Occupancy |
|---|---|---|---|---|---|
| Pb1 | *6g* | 0.73978 | 0.0 | 0.5 | 1.0 |
| Pb2 | *6g* | 0.0 | 0.38701 | 0.5 | 1.0 |
| Te1 | *4f* | 0.6667 | 0.3333 | 0.60511 | 1.0 |
| Te2 | *6h* | 0.33775 | 0.16888 | 0.25 | 1.0 |
| Ni1 | *4f* | 0.6667 | 0.3333 | 0.39826 | 1.0 |
| Ni2 | *6h* | 0.16476 | -0.16476 | 0.25 | 1.0 |
| Ni3 | *6h* | 0.51023 | 0.48977 | 0.25 | 1.0 |
| Ni4 | *2b* | 0.0 | 0.0 | 0.25 | 1.0 |
| O1 | *12i* | 0.3357 | 0.0084 | 0.1716 | 1.0 |
| O2 | *12i* | 0.1779 | 0.0123 | 0.3281 | 1.0 |
| O3 | *12i* | 0.4843 | 0.3271 | 0.1697 | 1.0 |
| O4 | *12i* | 0.1704 | 0.3299 | 0.1644 | 1.0 |
| O5 | *12i* | 0.6537 | 0.1789 | 0.5025 | 1.0 |

**Table S II: The details of magnetic couplings in the honeycomb like layer.**

| Magnetic coupling path | Bond-length (Å) | Bond-angle (°) | Expected coupling type |
|---|---|---|---|
| Ni2-O2-Ni4 | 3.0 | 92.4 | FM |
| Ni2-O1-Ni3 | 2.91 | 90.7 | FM |
| Ni2-O4-Ni3 | | 91.1 | |
| Ni1-O3-Ni3 | 3.51 | 122.2 | AFM |

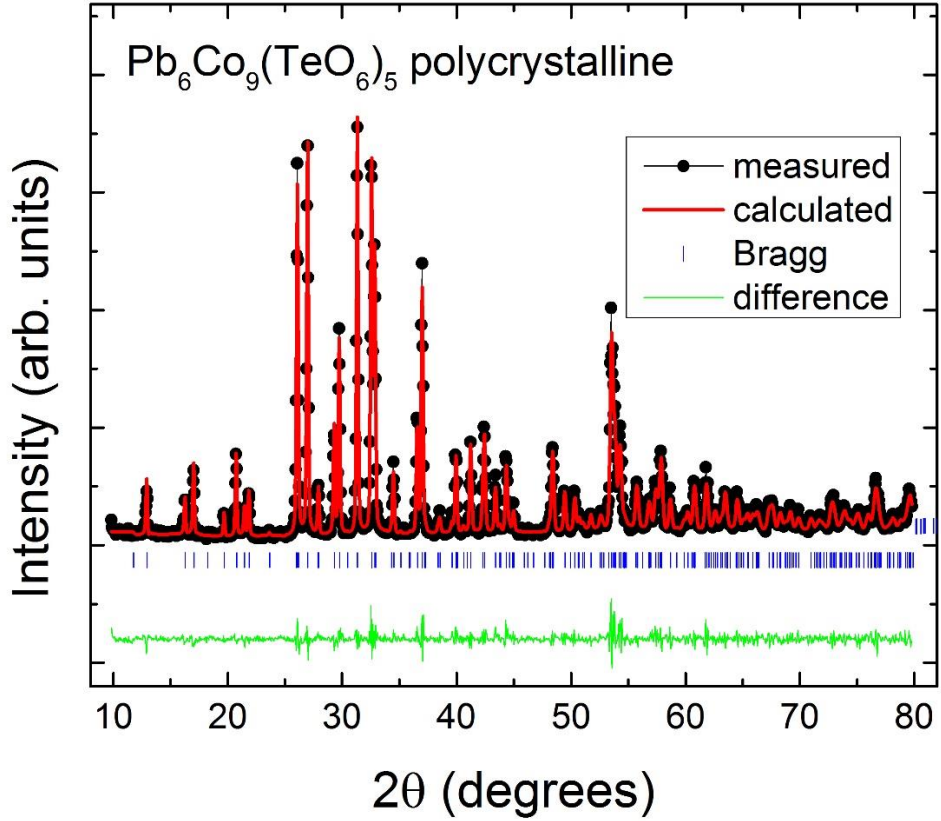

**Figure S2**: Rietveld refinement on the powder XRD of $Pb_6Co_9(TeO_6)_5$ samples.

**Table S III: The Reitveld refinement parameters of $Pb_6Ni_9(TeO_6)_5$ and $Pb_6Co_9(TeO_6)_5$ samples. It can be seen that the goodness of fit to the Co-sample is better than that of Ni-sample.**

| Refinement Parameters | $Pb_6Ni_9(TeO_6)_5$ | $Pb_6Co_9(TeO_6)_5$ |
|---|---|---|
| $R_p$ | 24.3 % | 16.7 % |
| $R_{wp}$ | 22.3 % | 18.2 % |
| $R_{exp}$ | 8.55 % | 13.5 % |
| $\chi^2$ | 6.83 | 1.81 |

**Table S IV: The parameters obtained from the magnetic data analysis of $Pb_6Ni_9(TeO_6)_5$ and $Pb_6Co_9(TeO_6)_5$ samples.**

| sample | C | $\mu_{eff}$ | $\theta_{CW}$ | $T_N$ |
|---|---|---|---|---|
| $Pb_6Ni_9(TeO_6)_5$ | 1.26 | 3.18 $\mu_B$ | -30 K | 25 K |
| $Pb_6Co_9(TeO_6)_5$ | 3.16 | 5.02 $\mu_B$ | -28 K | 26 K |

16